\documentclass[
journal=ancac3,manuscript=article]{achemso}
\usepackage[version=3]{mhchem} 
\usepackage{graphicx}  
\usepackage{amsmath} 
\usepackage{dcolumn}   
\usepackage{bm}        
\usepackage{amssymb}   
\usepackage{epstopdf}
\usepackage{xcolor} 
\usepackage{hyperref}
\usepackage[normalem]{ulem}
\author{Gaurav Pande}
\affiliation{Department of Physics, National Taiwan University, Taipei, Taiwan}
\alsoaffiliation{Nano Science and Technology Program, Taiwan International Graduate Program, Academia Sinica, Taiwan}
\author{Jyun-Yan Siao}
\affiliation{Department of Physics, National Taiwan University, Taipei, Taiwan}
\author{Wei-Liang Chen}
\affiliation{Center for Condensed Matter Sciences, National Taiwan University, Taipei, Taiwan}
\author{Chien-Ju Lee}
\affiliation{Department of Electrophysics, National Chiao Tung University, Hsinchu, Taiwan}
\author{Raman Sankar}
\affiliation{Institute of Physics, Academia Sinica, Taiwan}
\alsoaffiliation{Center for Condensed Matter Sciences, National Taiwan University, Taipei, Taiwan}
\author{Yu-Ming Chang}
\affiliation{Center for Condensed Matter Sciences, National Taiwan University, Taipei, Taiwan}
\author{Chii-Dong Chen}
\affiliation{Institute of Physics, Academia Sinica, Taiwan}
\author{Wen-Hao Chang}
\affiliation{Department of Electrophysics, National Chiao Tung University, Hsinchu, Taiwan}
\author{Fang-Cheng Chou}
\affiliation{Center for Condensed Matter Sciences, National Taiwan University, Taipei, Taiwan}
\alsoaffiliation{Taiwan Consortium of Emergent Crystalline Materials ($\mathit{TCECM}$), $\mathit{MOST}$, Taiwan}
\author{Minn-Tsong Lin} \email{mtlin@phys.ntu.edu.tw}
\affiliation{Department of Physics, National Taiwan University, Taipei, Taiwan}
\altaffiliation{Research Center for Applied Sciences, Academia Sinica, Taiwan}
\alsoaffiliation{Institute of Atomic and Molecular Sciences, Academia Sinica, Taiwan}

\title[An \textsf{achemso} demo]
  {Ultralow Schottky Barriers in h-BN Encapsulated Monolayer WSe$_2$ Tunnel Field-Effect Transistors}

\begin{document}

 \begin{tocentry}

\includegraphics[width=0.97\columnwidth]{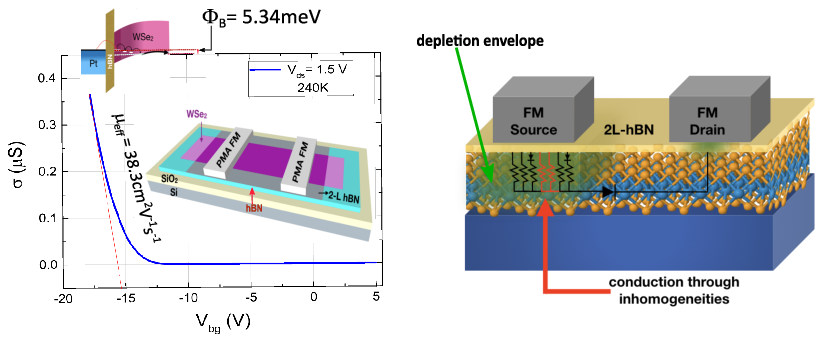}




  \end{tocentry}

\begin{abstract}

To explore the potential of field-effect transistors (FETs) based on monolayers of the two-dimensional semiconducting channel(SC) for spintronics, the two most important issues are to ensure the formation of variable low resistive tunnel ferromagnetic contacts(FC), and to preserve intrinsic properties of the SC during fabrication. Large Schottky barriers lead to the formation of high resistive contacts and methods adopted to control the barriers often alter the intrinsic properties of the SC. This work aims at addressing both issues in fully encapsulated monolayer WSe$_2$ FETs by using bi-layer h-BN as a tunnel barrier at the FC/SC interface. We investigate the electrical transport in monolayer WSe$_2$ FETs with current-in-plane geometry that yields hole mobilities $\sim$ 38.3 $cm^{2}V^{-1}s^{-1}$ at 240 K and On/Off ratios of the order of 10$^7$, limited by the contact regions. We have achieved ultralow effective Schottky barrier ($\sim$ 5.34 meV) with encapsulated tunneling device as opposed to a non-encapsulated device in which the barrier heights are considerably higher. These observations provide an insight into the electrical behavior of the FC/h-BN/SC/h-BN heterostructures and such control over the barrier heights opens up the possibilities for WSe$_2$-based spintronic devices.
\end{abstract}
\vspace{0.5cm}

\vspace{0.5cm}

\section{Introduction}
A large number of two-dimensional materials, including graphene, hexagonal boron nitride (h-BN), transition metal dichalcogenides (TMDCs) like MoS$_2$ and WSe$_2$, metal oxides (MxO$_y$), black phosphorene (b-P), etc, provide a wide range of properties and numerous applications feasible. Due to potential in a wide range of applications in nano-scale devices such as logic circuits \cite{lin_ambipolar_2014}, FETs \cite{pradhan_field-effect_2014,li_fabrication_2012}, LEDs \cite{withers_light-emitting_2015}, ultra-thin flexible devices \cite{he_fabrication_2012}, photodetectors \cite{wang_ultrasensitive_2015,nguyen_highly_2018} and so forth, two-dimensional materials such as atomically thin transition metal dichalcogenides(TMDCs) have been of significant interest to nano-electronics. In particular, single-layer TMDCs can be used in field effect transistors (FETs) as a high-mobility semiconductor channel \cite{radisavljevic_mobility_2013}, resulting in significant On/Off current ratios ($I_{\rm on}/I_{\rm off}>10^8$) \cite{radisavljevic_single-layer_2011} and reduced power dissipation \cite{schwierz_graphene_2010}. The coupled spin-valley physics \cite{xiao_coupled_2012} in TMDCs has also attracted broad attention in recent years as it provides a new opportunity for spin(opto-) valleytronic  applications \cite{chen_optical_2017} owing to the exotic effects arising from breaking of the inversion symmetry in monolayers and strong spin-valley coupling. Besides these features, the atomically thin nature of 2D TMDCs is significant, in the sense that it allows for effective electrostatics to easily control carrier density in the channel by gate voltages since the thickness of the channel falls below the charge depletion region on the metal/TMDC interface. One of the excellent starting points for the development of high performance digital electronic devices therefore is to comprehend the behavior of the metal in contact with TMDCs \cite{mott_note_1938}. Schottky contacts with low barrier heights ($\Phi_{B}$) and low reverse leakage currents are key requirements in order to produce TMDC-based devices like FETs that simultaneously preserves the intrinsic characteristics of the channel. This idea can also be extended to spin injection in these materials as long as there is a ferromagnetic(FM) contact \cite{chen_control_2013} with a tunnel barrier such that the height of the barrier is low enough to enable optimization of values for contact and channel resistance. This is called impedance matching phenomenon \cite{allain_electrical_2015} and is a key element for injection and detection of spin in semiconductors.

Analysis of Schottky transistor's electrical characteristics at room temperature alone cannot provide a complete understanding of the mechanism of conduction and formation of barriers. However, temperature-dependent electrical behaviour investigation may provide a detailed mechanism for current transport. To our knowledge, very few research in this direction of contact engineering for future lateral spin FET functionality has been made using FM contacts to monolayer TMDCs with an efficient tunnel barrier \cite{chen_control_2013,cui_low-temperature_2017,ghiasi_bilayer_2018}; however contacts with in-plane magnetization might be a bottleneck for fully electrical spin transport proposed in the monolayers in that case, since MLs of TMDCs prefer spin-polarization in z-direction originating from Ising Spin orbit coupling due to breaking of mirror symmetry of the plane perpendicular to 2d lattice plane \cite{zhou2019spin}, despite the gate tunability of the barrier heights. Subsequent efforts were made to combine WS$_2$ with perpendicular anisotropic magnets with Al$_2$O$_3$ barriers but even in that case, the Schottky barrier heights are considerably higher with the best value of 147 meV without any apparent gate tunability\cite{chen_optical_2017}.

This work focusses on the measurement and understanding of the gate-dependent and temperature-dependent charge transport properties in tunnel field-effect transistors formed by monolayer of WSe$_2$ crystals with ferromagnetic metal contacts with perpendicular magnetic anisotropy(PMA). Our findings show that, in combination with a thin layer of h-BN on top of ML WSe$_2$ along with thick h-BN from bottom, not only allows for ultra low effective barrier heights, but also allows this barrier height to be tuned with the gate voltages opening the possibility for modulation of these heights to yield comparable values for contact and channel resistance in future for impedance matching and spin injection. The strategy of encapsulating allows for probing the intrinsic transport properties in the monolayer SC while giving out the minimum effective Schottky barrier height as low as 5.34 meV by using two-dimensional thermionic emission model \cite{anwar_effects_1999} free from Fermi level pinning (FLP) effects. This tunability is manifestly lost in the un-encapsulated device. For simplicity, here onwards we will refer the contact material as Pt, the part of the superstructure which is in immediate contact with bilayer h-BN and ML WSe$_2$.
   
\begin{figure}[ptb]
\begin{center}
\includegraphics[width=1.0\columnwidth ]{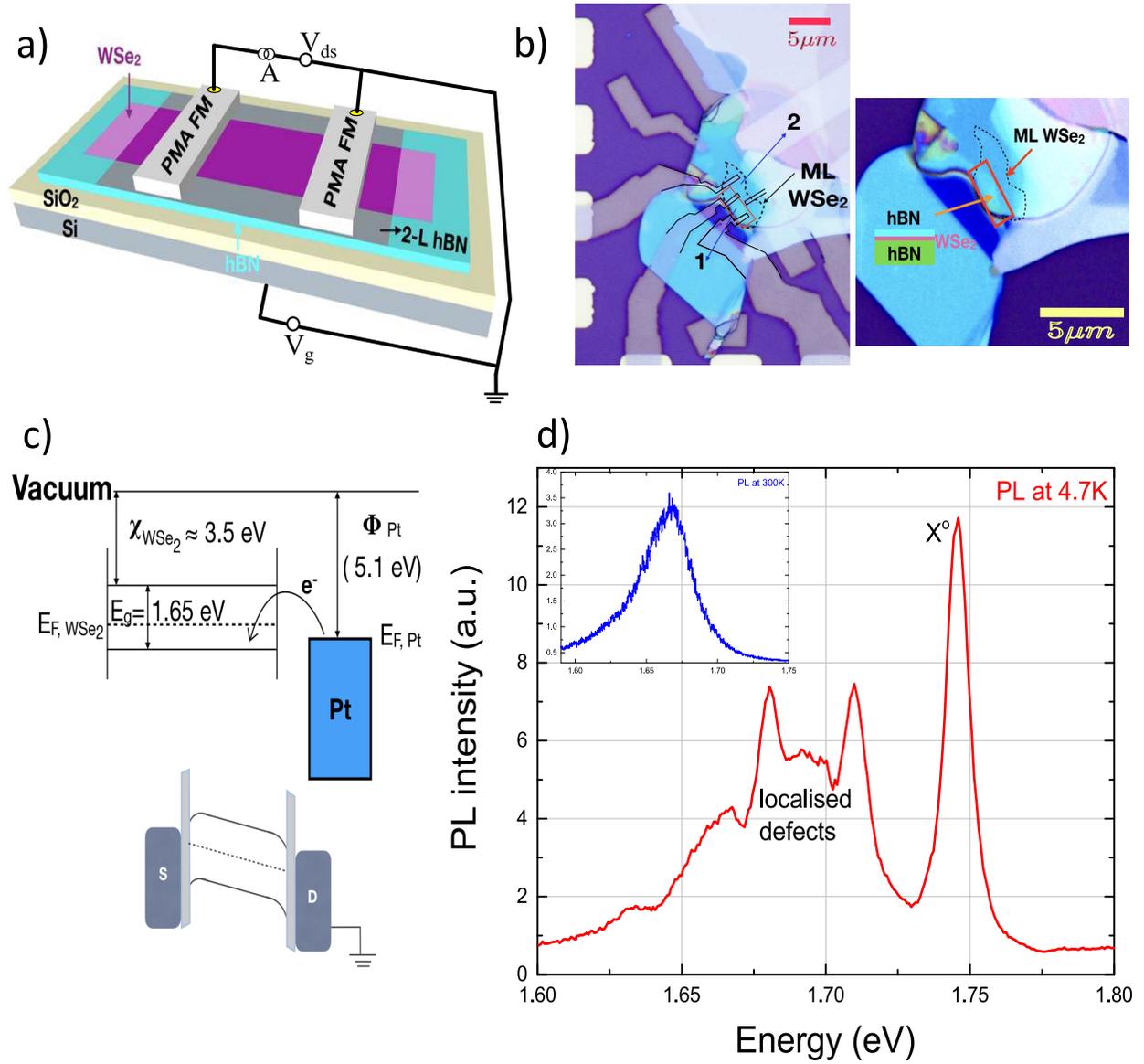}
\caption{(a) Schematic of the h-BN encapsulated monolayer WSe$_2$ tunneling device with ferromagnetic contacts (b) Optical microscope image of the device. The rectangular part(red) represents the encapsulated structure; Optical image of the encapsulated sample before defining contacts. (c) (Top)The energy level diagram for monolayer WSe$_2$ with respect to the immediate contact material platinum; (Bottom)The schematic of drain-source current under the forward bias condition under finite bias and over threshold gate voltage. Note that the majority charge carriers are holes in our devices. The band bending around the ferromagnetic contacts is not scaled. (d) Photoluminescence (PL) spectrograph for monolayer WSe$_2$ at 4.7K (X$^o$ represents the neutral exciton peak);(inset) Room temperature PL spectra for the same single layer of WSe$_2$ shows the single signature peak at 1.67 eV for collective excitations in a monolayer.
}
\label{fig:schematic}
\end{center}
\end{figure}

\section{Results and Discussion}

{\bf Device fabrication}

Monolayer (ML) WSe$_2$ and thin h-BN are obtained by mechanical exfoliation from bulk 2H- WSe$_2$ (see supplementary data for bulk crystal growth details) and h-BN crystals (2D Semiconductors Inc.) by polydimethylsiloxane(PDMS) based method followed by subsequent dry transfer with the help of a micro-manipulator onto boron doped SiO$_2$/Si (100) substrate (300 nm thick SiO$_2$). Initially, the layer thickness is determined by the optical contrasts. This procedure has been followed to create two devices. One assembly forms 2L-hBN/WSe$_2$/hBN stack (D1) while the other WSe$_2$/hBN stack (D2) by deliberately misplacing the thin h-BN flake and the ML WSe$_2$ flake to yield one encapsulated area and one non-encapsulated, followed by annealing in Ar/H$_2$ atmosphere at 423K for 3 hrs. Fig. 1a shows the schematic of the tunnel FET device. In an effort to reduce the inhomogeneity between the devices, both the studied transistors are fabricated simultaneously on a single substrate. The Si is used as a back gate to tune the carrier density in the channel. WSe$_2$ is confirmed to be a single layer by confocal photoluminescence(PL) measurements as shown in Fig. 1d (see supplementary data for details). The high quality of encapsulated SC with thin h-BN is reflected in the high neutral exciton X$^o$ signal at low temperature PL depicted in Fig. 1d compared to the defect peaks as opposed to an un-encapsulated sample lying bare on SiO$_2$/Si substrate shown in Fig. S3 (see supplementary data) obtained from the same bulk. It is important to note that the WSe$_2$ beneath the thin h-BN does not come in contact with polymers during the fabrication due to full encapsulation of the flake. ML WSe$_2$ field effect transistor devices are fabricated by patterning MMA/PMMA [(Methyl Methacrylate)/poly-(Methyl Methacrylate)] resists using e-beam lithography. The metallic ferromagnetic electrodes consisting of a superstructure formed by Co/Pt with Pt(15\AA) as the immediate contact material on top of both D1 and D2 are deposited inside an ultrahigh vacuum (UHV) chamber with base pressure $\sim$10$^{-9}$ Torr. The magnetic behavior is confirmed by polar magneto-optical Kerr effect (see Fig. S4) with external magnetic field perpendicular to the sample surface. The channel length of both the devices is 4 $\mu$m and 3.5 $\mu$m respectively and the width of the flakes is generally between 2 and 2.5 $\mu$m. The dielectric screening by h-BN on ML WSe$_2$ not only helps with device mobility but also protect the channel from environmental effects and degradation. The optical microscope image shown in Fig. 1b illustrates the final device geometry for single layer WSe$_2$ FETs. Fig. 1c illustrates the energy level diagram for monolayer WSe$_2$ with respect to the immediate contact material platinum to the channel via thin h-BN barrier(not shown). Current-Voltage (I-V) measurements are performed by applying a DC voltage across two electrodes and registering the current response in both devices.

\begin{figure}[ptb]
\begin{center}
\includegraphics[width=1.0\columnwidth]{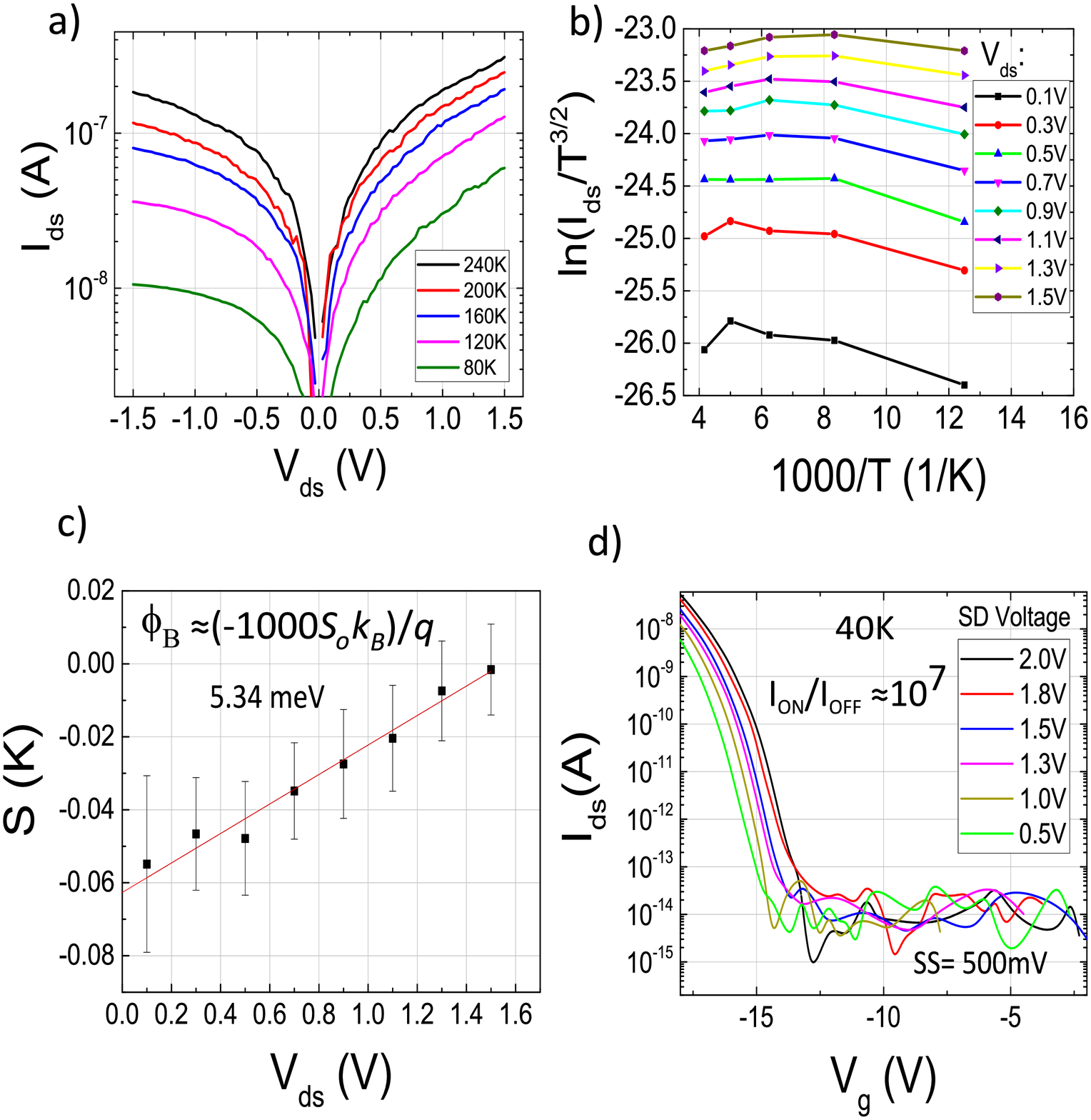}
\caption{Charge transport in encapsulated ML WSe$_2$ device at V${_g}$= -18V . (a) I-V curves for the monolayer WSe$_2$ FET device with FM contacting to WSe$_2$ via a tunnel barrier formed by 2-layer h-BN from T=80K to 240K (b) Arrhenius plot $ln(I_{ds}/T^{3/2})$ vs 1000/T at different drain-source voltages V$_{ds}$ in the thermionic emission temperature range (c) Extraction of Schottky barrier height $\Phi _{B}$  by taking the y-intercept value. Here each data point represents the slope obtained from the Arrhenius plot in (b) under a specific V$_{ds}$ (d) Current I$_{ds}$ plotted on a logarithmic scale as extracted from a single layer WSe$_2$ FET at T=40K and as a function of the gate voltage V$_g$ for several values of the voltage V$_{ds}$ between drain and source contacts. Notice that the ON/OFF ratio approaches $10^7$ and sub-threshold swing SS$\sim$ 500mV per decade.}
\label{fig:barrierheight}%
\end{center}
\end{figure}

{\bf DC bias electrical transport and Schottky analysis}

We now examine the two probe I-V curves as a function of temperature in order to better comprehend the essence of the Schottky barrier in these devices. First, we discuss the I-V characteristics for tunnel device with Pt/h-BN contacts to ML WSe$_2$. Fig. 2a displays I-V curves on a logarithmic scale for several temperatures falling into thermionic emission regime. The measured current response increases with the increase in temperature. The figure demonstrates a rectifying behavior from symmetric electrodes especially at 80K. Experimentally, even when designing symmetric electrodes, there is some difference in the widths of the electrodes on the order of 'nm'. In fact, it is quite difficult to make them ideally symmetric. Hence, we believe that at low temperatures, because of the slight asymmetry of the electrodes, one of the electrodes starts behaving more resistive than the other one. Therefore, the nearly symmetrical rectifying behavior in the I$_{ds}$-V$_{ds}$ at low temperatures is due to this back to back Schottky-diode-like structures. A typical two-probe I-V curve taken at a wide temperature range is displayed in Fig. S2a (see supplementary data) for this device with channel length, L, of 4 $\mu$m. Please note that our device channel length is $\sim$4$\mu$m while the contact widths are $\sim$1$\mu$m. For a constant channel length, since the tunneling/ (Schottky Barrier) changes exponentially with voltage, increasing above some voltage results in a very small change of voltage on the contacts, and most of it falling on the channel, and if the channel is larger comparative to contacts, leading to a (nearly) linear IV behavior. Two probe conductance measured at 240K as a function of V$_g$ is shown in Fig. S2b(see supplementary data) and exhibits decreasing conductance as V$_g$ is swept towards positive voltages. In fact, this characteristic behaviour is seen on both devices measured throughout the wide temperature range and indicates that the devices are intrinsically p-type as fabricated. It reflects that the Fermi level lies deep within the band gap close to valence band of WSe$_2$ which is expected of a very high work function metal contact material like platinum here. The back gate voltage V$_g$ applied to Si substrate results in 500mV sub-threshold slope for every decade of current change shown by Fig. 2d at 40K (for 300 nm SiO$_2$ dielectric). Electrical transport across a Schottky contact into a semiconductor has been conventionally described by the 3D thermionic emission equation\cite{bhuiyan_new_1988}. 


The same format of the equation is modified into a slightly different equation when we consider devices which falls in the regime of thin 2D materials as is our case.  In such a situation, the drain-source current I$_{DS}$ can be defined by a 2D thermionic emission equation \cite{anwar_effects_1999} instead. This equation has a form of a reduced power law in temperature T$^{3/2}$ for a two dimensional transport channel and is given by:

\begin{equation}
I_{DS}=A_{2D}^{*}ST^{3/2}\exp\left [-\frac{q}{\mathit{k_{B}}T}\left ( \Phi _{B}-\frac{V_{DS}}{n} \right )  \right ]
\end{equation}

where $A^{\ast}$ is the two-dimensional equivalent Richardson constant, $S$ is the contact area of the junction, $q$ is the elementary charge, $\Phi _{B}$ is the Schottky barrier height, $n$ is the ideality factor, $\mathit{k_{B}}$ is the Boltzmann constant and $V_{DS}$ is the drain-source bias.

To determine $\Phi_{B}$, $ln(I_{ds}/T^{3/2})$ is plotted against $1000/T$ for various V$_{DS}$ biases as shown in Fig. 2b. The data is linear at each bias in the thermionic emission regime and the slope(S) is subsequently plotted in Fig. 2c as a function of drain-source bias. The y-intercept of Fig. 2c, denoted by S$_o$ in the formula mentioned on the figure, yields $\Phi_{B}$ according to $S_{o}=-q\Phi _{B}/1000\mathit{k_{B}}$. For ML WSe$_2$ transistors D2 with direct Pt contacts, $\Phi_{B}$ is found to be $\sim$ 239 meV (Fig. 3b) and shows hardly any modulation with the change of gate voltages. On the other hand, the tunnel barrier encapsulated device D1 attains a good gate tunability (Fig. 3a) and shows an effective Schottky barrier height as low as 5.34 meV as shown in Fig. 2c. It is worth mentioning that the true estimate of the Schottky barrier height comes from the flat band deviation of effective Schottky values from a linear region and since the device without encapsulation D2 does not show any signs of tunability (see Fig. 3b), it is very hard to pin point the specific value of the barrier height in that case. Such analysis,on the other hand, seemed possible with the encapsulated device but the interference from leakage possibly via SiO$_2$ for a broader range of gate voltages has not enabled us to make that conclusion decisively. Even so, the lower bound for the 'true' barrier height could be inferred as 32.5 meV as shown in Fig. 3a since this value scales with the threshold gate voltage of the transistor within the temperature range where thermionic emission model is valid. In addition, since the device can be tuned for the Schottky heights, a transition from thermionic emission regime possibly into the tunneling dominant regime is achieved pushing this effective value down to an ultralow of 5.34 meV.

The D2 corresponds to direct Schottky contact by Pt to the ML WSe$_2$ flake. Temperature dependent I-V characteristics shows an asymmetry on either side of the bias and the transistor behaves more like a Schottky diode shown in Fig. S1d (see supplementary data). This reflects the dominance of transport assisted by thermionic emission and is consistent with the previous results \cite{jariwala_band-like_2013,yamaguchi2014tunneling,moriya2014large} for direct normal metal (NM)-SC junction. On the other hand for D1, as has been shown in Fig. 2a, an almost symmetric behaviour in I$_{DS}$-V$_{DS}$ plot around zero bias could be seen, suggesting dominance of tunnel transport over thermionic emission\cite{omar2017graphene}. Thus, instead of increasing two-probe resistance, insertion of a 2L-hBN barrier reduces the device resistance, especially at low drain-source biases which could be quite relevant for all practical purposes of electrical spin injection/detection performed generally at low-biases \cite{liang2017electrical}. The two terminal resistances at 10K in the low bias regime exceed 500 M$\Omega$ for tunnel device D1 shown in Fig. S2(inset). This can be confirmed quantitatively by examining the 2D Arrhenius behaviour using eq. 1 and extracting $\Phi_{B}$. Fig. 2b shows Arrhenius plot at different drain source bias voltages. Fig. 2c displays resulting Schottky barrier height extracted from extrapolation of slopes in Fig. 2b as 5.34 meV as discussed above. This is a dramatic decrease, of almost 98\%, in $\Phi_{B}$  from no barrier to 2L-hBN barrier (Fig. 3a and 3b). Notably, a significant fact is that the barrier height, which is related to the contact resistance, can be controlled by inserting h-BN barriers. Fig. 2d shows the sub-threshold slope curve for estimation of On/Off ratio for the tunnel FET. The data shown for 40K in Fig. 2d as a function of gate voltages shows the best value for this ratio is of the order of 10$^7$ which is comparable to previous high performance devices made from WSe$_2$ monolayer as SC with normal metal contacts\cite{liu2013high}.

\begin{figure}[ptb]
\includegraphics[width=1.0\columnwidth]{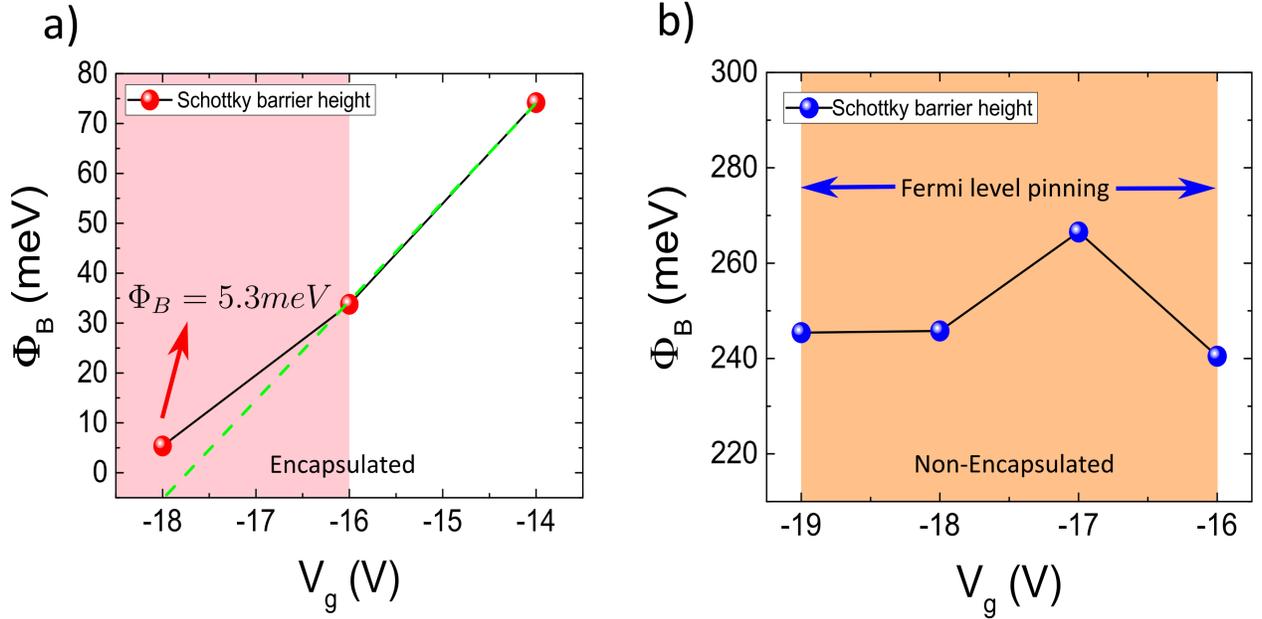}
\caption{(a) Back-gate voltage dependent effective Schottky barrier heights $\Phi_{B}$ for 2-layer hBN tunnel barrier device. The deviation from the linear response at V$_g$ = -16V (green dotted line) should define the flat band voltage (V$_F$) and close to the real $\Phi_{B}$ for Pt/hBN on monolayer WSe$_2$. The shaded pink region suggests a crossover from the thermionic regime into pure tunneling regime. (b) Back-gate voltage dependent effective Schottky barrier heights $\Phi_{B}$ for not encapsulated ML WSe$_2$ device with direct metal-semiconductor junction. Notice the high values and non-tunability of the barrier heights attributed to FLP effects (orange area).
}
\label{fig:Arrhenius}%
\end{figure}

The Schottky barrier height is dependent solely on the work functions of the metal and SC for metals-SC junctions without interface states \cite{donaldjneamen2012}. A research on MoS$_2$ Schottky contacts showed that the measured height of the Schottky barrier depends linearly on work functions \cite{das2012high}. However, it is an open issue whether or not interfacial states also play a part. It is known that such states can dramatically affect $\Phi_{B}$ through Fermi level pinning\cite{hughes_metal-gallium_1982, lagasse_gate-tunable_nodate} as exhibited in our device D2. For a high density of interfacial states (DIS), $\Phi_{B}$ can be completely determined by the interfacial states, independent of the metal work function. The gate dependence of the barrier provides some insight on the role of interfacial states. In the conventional theory of Schottky barriers for semiconductors,\cite{cowley1965surface} both work functions and interface states play a role in determining the barrier height with relative importance depending on the DIS. Interestingly, in the two limits of DIS = 0 (work function model) and DIS $\rightarrow \infty $ (FLP) the Schottky barrier is independent of the electron density. However, in the intermediate regime of DIS, the Schottky barrier height decreases with increasing electron density. Since the same trend is observed in our ML WSe$_2$ transistor D1, it provides an evidence that work functions and interfacial states both play a role in determining the barrier height. Therefore, the data presented here for ML WSe$_2$, which also demonstrates a lowering of the Schottky barrier height with insertion of 2L-hBN barrier, begs an explanation for the case of intermediate DIS and suggests that interfacial states are important for a full understanding of Schottky contacts to TMDCs. At the same time, the likelihood of transport being affected by intrinsic defects in the bulk and fabrication related details, could also be an open possibility. A detailed model explaining that, in addition to the experimental data and technique to probe the interface directly, is worthy of further investigation however out of the scope of this work.

I-V measurements combined with the gate dependence provide additional insight into the impacts of h-BN barriers on $\Phi_{B}$. For Pt directly contacted to ML WSe$_2$ without the h-BN barrier, the Schottky barrier height does not decrease considerably and varies approximately from 239 $\pm$ 5.9 to 258 $\pm$ 1.6 meV at various gate voltages as shown in Fig. 3b (blue dots). With the insertion of 2L-hBN barrier, $\Phi_{B}$ varies from 5.34 $\pm$ 0.08 meV at V$_g$ = -18V to 72 $\pm$ 2.6 meV at V$_g$ = -14 V. In the measured range, the direct Pt contact Schottky barrier height is never less than for Pt with 2L-hBN. A simple linear extrapolation indicates that, for no barrier, $\Phi_{B}$ is approximately 258 meV for gate voltages above -14 V. On the other hand, with the insertion of 2L-hBN $\Phi_{B}$ is relatively low at all measured gates and goes to ultra-low 5.34 meV at V$_g$ = -18 V. Thus, both back gate and 2L-hBN provide a wide parameter space to control the Schottky barrier height in ML WSe$_2$ tunnel FET device.

{\bf Discussion}

\begin{figure}[ptb]
\includegraphics[width=1.0\columnwidth]{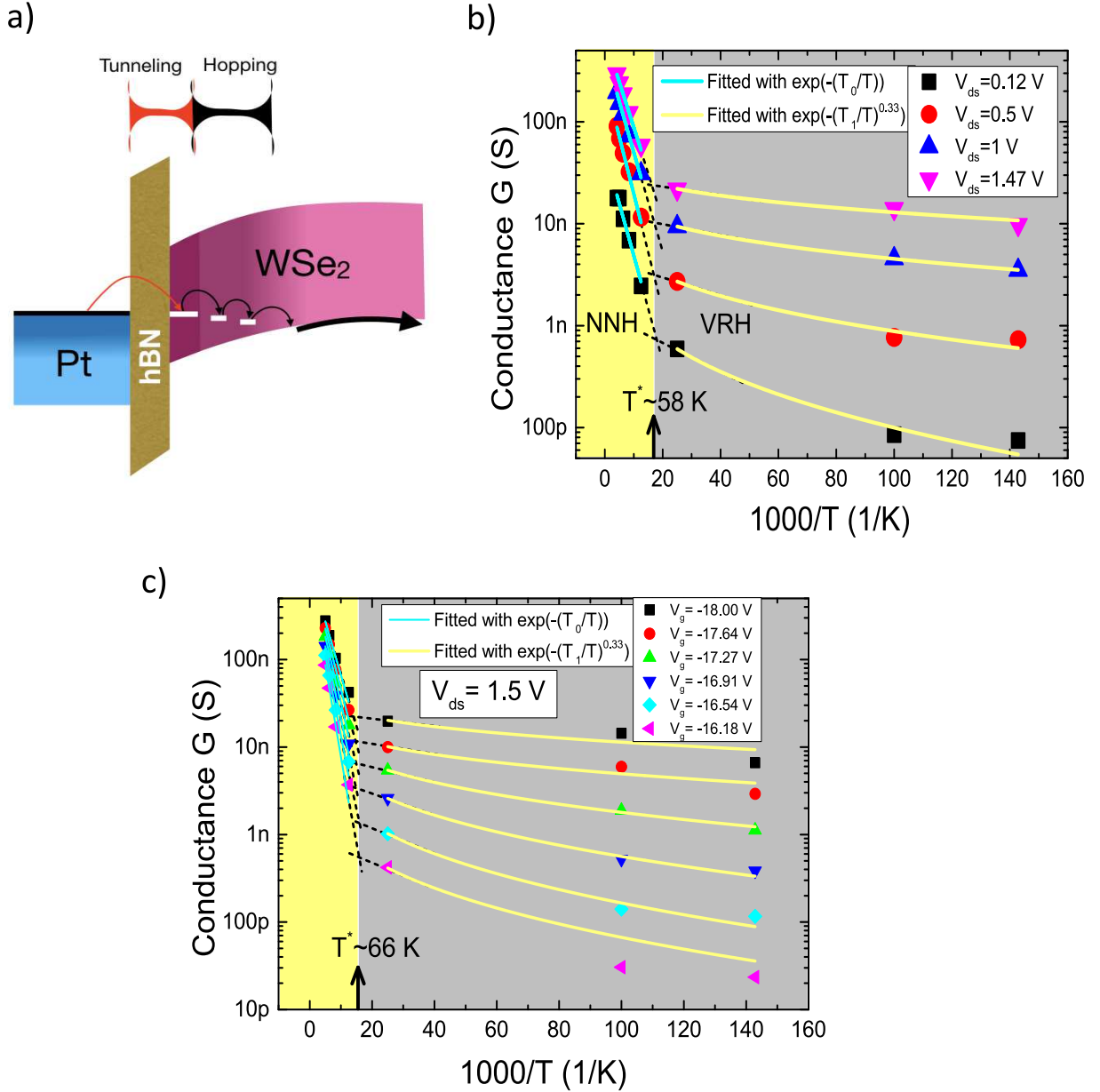}
\caption{ Evidence for hopping transport in Pt-hBN-WSe$_2$ contact region (a) Band diagram for the Schottky contact region for ML WSe$_2$ device. The device can be separated into three locales. The direct tunneling locale comprises of the h-BN tunnel barrier and one part of Schottky contact taken together. The second locale is in the tail portion of depletion layer where holes(electrons) transport in a hopping mannerism. The third locale is where holes(electrons) either transport in the WSe$_2$ channel by a hopping or transport in the WSe$_2$ valence band(conduction band), contingent upon the carrier density. (b) Arrhenius plot of the temperature dependent conductance G(symbols) at various V$_{ds}$ at V$_g$ = -18V and the fitting results by the different hopping models (blue and yellow lines). Two hopping regimes are clearly separated by $T^{\ast}$ (vertical line) corresponding to a temperature of $\sim$58K.(c) Arrhenius plot of G for different values of V$_g$. Solid lines are fits to the data showing two distinct conduction behaviour for various regions of T and V$_g$ (charge density).
} 
\label{fig:tunneling}%
\end{figure}

To reconcile the case for efficient tunneling in our encapsulated device, one has to be careful to see the Arrhenius plots over a broad temperature range . To clarify this point, let us focus on the temperature dependence of conductance which serves as a fingerprint for the electronic hopping processes that are involved in the transport as shown in the schematic in Fig. 4a. In WSe$_2$ system, it is reported that the selenium vacancies can introduce localized donor states inside the band gap \cite{stanford2017high}. Fig. 4b shows the Arrhenius plot for T-dependence conductance at different V$_{ds}$ at a constant gate voltage. It could be seen from temperature variation of G with different V$_{ds}$ that the charge transport phenomena can be described by two different mechanisms, known as activation transport models, in respective high and low temperature regimes, with a characteristic threshold at around $T^{\ast }$ = 58K. For this $T> T^{\ast }$, carriers with a conductivity varying like $G\sim \exp (-T_{0}/T)$ are dominated by the nearest-neighbor hopping (NNH) mechanism. For $T< T^{\ast }$, a 2D variable-range hopping (VRH) mechanism equation with much weaker temperature dependence according to $G\sim \exp (-T_{1}/T)^{0.33}$ can instead be used to fit the conductance. In many low-dimensional systems, such T-dependence has been observed in accordance with Mott variable-range hopping theory \cite{doi:10.1080/14786437208226975} and is a signature for transport via localized states driving the system into a strongly localized regime \cite{qiu_hopping_2013,han_electron_2010,yu_variable_2001}. In Fig. 4c, as a different set of data to verify the fitting models used above, which shows temperature variation of G with different V$_g$s instead, for the same encapsulated sample D1, the slope of Arrhenius plot changes around the similar $T^{\ast }$, implying the same crossover of conduction mechanisms. The two temperature regions for the different hopping regimes are even more pronounced at small drain-source bias $ V_{ds}$ of 0.12V where the probability of contact resistance dominating the total resistance is quite high. This emphasizes that the dominant transport mechanism is hopping through the contact regions over the WSe$_2$ channel itself. 

In this instance, we put forward an explanation that the contact region may be constituted by three different zones as depicted schematically in Fig. 4a: (i) the dielectric insulating h-BN tunnel barrier, (ii) the strongly depleted zone (dark pink) underneath h-BN playing the role of an additional composite tunnel barrier and (iii)the rear part of the depletion zone (smeared pink) of the ML WSe$_2$ where the hopping mechanism guarantees electronic conduction. If we pay close attention to the inhomogeneous channel composed of two components: the depletion part below h-BN and a semi-infinite section of a channel, electron hopping can profit from the electric field \cite{lu_electrical_2014} or thermal activation energy \cite{lu_spin-polarized_2009}through localized states whenever there is a rise in either the bias or temperature. Particularly for the variable range hopping procedure and can effectively reduce resistance R$_ {iii}$. This fact could be investigated in more details by experiments with higher number of probes which can shed more light precisely on this subject through an elaborate investigation of channel resistances. These results are extremely promising indicators in the direction of the possibility of spin injection and transport in encapsulated ML WSe$_2$. Uniform h-BN barriers can provide the contact resistance needed to alleviate the conductance mismatch problem and the resistance value can be easily controlled by gate voltages, enabling the tuning of $\Phi_{B}$ down to almost zero given the limits for gate leakage values (this might vary from device to device). This permits a minimization of depletion region for a highly conductive spin transport channel.

\section{Conclusions}

We have examined the properties of p-type single layer tunneling WSe$_2$ FET devices with ferromagnetic electrodes. In particular, using 2D thermionic emission analysis by current-voltage curves, we have measured the Schottky barrier height for directly contacted Pt electrodes to WSe$_2$ and has attributed the non-tunability to FLP effects. Remarkably, we have found that, with the insertion of a thin h-BN barrier between the Pt electrode and the WSe$_2$ flake encapsulated by thicker h-BN, the effective Schottky barrier height can be reduced to an ultra-low by as much as by 97.7$\%$. We have shown that the barrier height can be manipulated by both h-BN presence and back gate voltages. From systematic studies of the bias, temperature and back-gate voltage dependence of Schottky barrier heights, we surmise that in order to maintain a balance between interface tunneling resistance and channel resistance, which is mandatory for observation of two-terminal spin signals, the understanding and the behavior of hopping via localized states in the contact depletion region could play a central role. Elimination of metal/TMDC chemical interactions by the h-BN insertion layer along with full boron nitride encapsulation of the channel allows to better preserve the intrinsic properties of the 2D channel. The discussion presented here for hopping via h-BN tunnel barriers to 2D semiconductors allows satisfactorily to describe and reproduce their experimental electrical response within the Schottky-Mott limit. Such low barrier heights and the ability to optimize them encourages and opens up the avenues for future studies on purely electrical spin transport in MLs of WSe$_2$. These results promises to be important not only to those interested in WSe$_2$ in particular, but also to the larger community exploiting 2D materials for optoelectronic and spintronics applications.

\section{Methods}

Monolayer (ML) WSe$_2$ and thin h-BN are obtained by mechanical exfoliation from bulk 2H- WSe$_2$ grown by Chemical vapor transport(see supplementary data for bulk crystal growth details) and h-BN crystals (2D Semiconductors Inc.) by polydimethylsiloxane(PDMS) based method followed by subsequent dry transfer with the help of a micro-manipulator onto boron doped SiO$_2$/Si (100) substrate (300 nm thick SiO$_2$). Initially, the layer thickness is determined by the optical contrasts. Room temperature PL spectra were acquired with a home-built laser scanning confocal spectro-microscope with 532 nm laser excitation. The low temperature photoluminescence (PL) spectra of WSe$_2$ monolayers were measured by a micro-photoluminescence setup. The setup consist of a 3-axis positioner and a 100X objective lens (NA = 0.82) integrated in a cryogen-free cryostat with a base temperature of 4.2 K. A He-Ne laser with a wavelength of 632.8nm was used to excite the WSe$_2$. The charge-transport measurements have been performed in a four probe cryostat system by \textit{Lakeshore} varying temperature from 6.6 K to 280 K with a base pressure of 2 $\times$ 10$^{-6}$ mbar . For the back-gated two-terminal measurements as described in Fig. 2, we have used one of Keithley 2636B channel to apply the drain-source bias $V_{ds}$ and measure drain-source current $I_{ds}$, and used the other Keithley 2636B channel to apply the back-gate voltage $V_g$. The estimation of the mobility of our FET device with 2L h-BN tunnel junction have been performed by the $I_{DS}-V_G$ characterization. All measurements reported here use a voltage biased scheme.

{\bf Acknowledgements}

We would like to acknowledge Ting-Ching Chu and Ching-Chou Tsai for constant fruitful discussions. This work was supported by the Ministry of Science and Technology (MOST), Taiwan through the following grants: MOST 107-2119-M-002-006 and MOST 108-2112-M-001 -049 -MY2.

{\bf Supporting Information Available}

Supporting information contains details of bulk single crystal preparation, Room and low temperature PL measurements, details of electrical measurements, Polar MOKE measurement of Co/Pt layered FM superstructure at room temperature and optical images of the device constituents. This material is available online at \href{http://pubs.acs.org}{http://pubs.acs.org}.

\bibliographystyle{achemso}

\begin{mcitethebibliography}{39}
\providecommand*\natexlab[1]{#1}
\providecommand*\mciteSetBstSublistMode[1]{}
\providecommand*\mciteSetBstMaxWidthForm[2]{}
\providecommand*\mciteBstWouldAddEndPuncttrue
  {\def\EndOfBibitem{\unskip.}}
\providecommand*\mciteBstWouldAddEndPunctfalse
  {\let\EndOfBibitem\relax}
\providecommand*\mciteSetBstMidEndSepPunct[3]{}
\providecommand*\mciteSetBstSublistLabelBeginEnd[3]{}
\providecommand*\EndOfBibitem{}
\mciteSetBstSublistMode{f}
\mciteSetBstMaxWidthForm{subitem}{(\alph{mcitesubitemcount})}
\mciteSetBstSublistLabelBeginEnd
  {\mcitemaxwidthsubitemform\space}
  {\relax}
  {\relax}


\bibitem[Lin \latin{et~al.}(2014)Lin, Xu, Wang, Li, Yamamoto,
  Aparecido‐Ferreira, Li, Sun, Nakaharai, Jian, Ueno, and
  Tsukagoshi]{lin_ambipolar_2014}
Lin,~Y.-F.; Xu,~Y.; Wang,~S.-T.; Li,~S.-L.; Yamamoto,~M.;
  Aparecido‐Ferreira,~A.; Li,~W.; Sun,~H.; Nakaharai,~S.; Jian,~W.-B.;
  Ueno,~K.; Tsukagoshi,~K. Ambipolar {MoTe}$_2$ {Transistors} and {Their}
  {Applications} in {Logic} {Circuits}. \emph{Advanced Materials}
  \textbf{2014}, \emph{26}, 3263--3269\relax
\mciteBstWouldAddEndPuncttrue
\mciteSetBstMidEndSepPunct{\mcitedefaultmidpunct}
{\mcitedefaultendpunct}{\mcitedefaultseppunct}\relax
\EndOfBibitem
\bibitem[Pradhan \latin{et~al.}(2014)Pradhan, Rhodes, Feng, Xin, Memaran, Moon,
  Terrones, Terrones, and Balicas]{pradhan_field-effect_2014}
Pradhan,~N.~R.; Rhodes,~D.; Feng,~S.; Xin,~Y.; Memaran,~S.; Moon,~B.-H.;
  Terrones,~H.; Terrones,~M.; Balicas,~L. Field-{Effect} {Transistors} {Based}
  on {Few}-{Layered} α-{MoTe} $_{\textrm{2}}$. \emph{ACS Nano} \textbf{2014},
  \emph{8}, 5911--5920\relax
\mciteBstWouldAddEndPuncttrue
\mciteSetBstMidEndSepPunct{\mcitedefaultmidpunct}
{\mcitedefaultendpunct}{\mcitedefaultseppunct}\relax
\EndOfBibitem
\bibitem[Li \latin{et~al.}(2012)Li, Yin, He, Li, Huang, Lu, Fam, Tok, Zhang,
  and Zhang]{li_fabrication_2012}
Li,~H.; Yin,~Z.; He,~Q.; Li,~H.; Huang,~X.; Lu,~G.; Fam,~D. W.~H.; Tok,~A.
  I.~Y.; Zhang,~Q.; Zhang,~H. Fabrication of {Single}- and {Multilayer} {MoS}$_2$
  {Film}-{Based} {Field}-{Effect} {Transistors} for {Sensing} {NO} at {Room}
  {Temperature}. \emph{Small} \textbf{2012}, \emph{8}, 63--67\relax
\mciteBstWouldAddEndPuncttrue
\mciteSetBstMidEndSepPunct{\mcitedefaultmidpunct}
{\mcitedefaultendpunct}{\mcitedefaultseppunct}\relax
\EndOfBibitem
\bibitem[Withers \latin{et~al.}(2015)Withers, Del Pozo-Zamudio, Mishchenko,
  Rooney, Gholinia, Watanabe, Taniguchi, Haigh, Geim, Tartakovskii, and
  Novoselov]{withers_light-emitting_2015}
Withers,~F.; Del Pozo-Zamudio,~O.; Mishchenko,~A.; Rooney,~A.~P.; Gholinia,~A.;
  Watanabe,~K.; Taniguchi,~T.; Haigh,~S.~J.; Geim,~A.~K.; Tartakovskii,~A.~I.;
  Novoselov,~K.~S. Light-Emitting Diodes by Band-Structure Engineering in Van
  der {Waals} Heterostructures. \emph{Nature Materials} \textbf{2015},
  \emph{14}, 301--306\relax
\mciteBstWouldAddEndPuncttrue
\mciteSetBstMidEndSepPunct{\mcitedefaultmidpunct}
{\mcitedefaultendpunct}{\mcitedefaultseppunct}\relax
\EndOfBibitem
\bibitem[He \latin{et~al.}(2012)He, Zeng, Yin, Li, Wu, Huang, and
  Zhang]{he_fabrication_2012}
He,~Q.; Zeng,~Z.; Yin,~Z.; Li,~H.; Wu,~S.; Huang,~X.; Zhang,~H. Fabrication of
  {Flexible} {MoS}$_2$ {Thin}-{Film} {Transistor} {Arrays} for {Practical}
  {Gas}-{Sensing} {Applications}. \emph{Small} \textbf{2012}, \emph{8},
  2994--2999\relax
\mciteBstWouldAddEndPuncttrue
\mciteSetBstMidEndSepPunct{\mcitedefaultmidpunct}
{\mcitedefaultendpunct}{\mcitedefaultseppunct}\relax
\EndOfBibitem
\bibitem[Wang \latin{et~al.}(2015)Wang, Wang, Wang, Hu, Zhou, Guo, Huang, Sun,
  Shen, Lin, Tang, Liao, Jiang, Sun, Meng, Chen, Lu, and
  Chu]{wang_ultrasensitive_2015}
Wang,~X.; Wang,~P.; Wang,~J.; Hu,~W.; Zhou,~X.; Guo,~N.; Huang,~H.; Sun,~S.;
  Shen,~H.; Lin,~T.; Tang,~M.; Liao,~L.; Jiang,~A.; Sun,~J.; Meng,~X.;
  Chen,~X.; Lu,~W.; Chu,~J. Ultrasensitive and {Broadband} {MoS}$_2$
  {Photodetector} {Driven} by {Ferroelectrics}. \emph{Advanced Materials}
  \textbf{2015}, \emph{27}, 6575--6581\relax
\mciteBstWouldAddEndPuncttrue
\mciteSetBstMidEndSepPunct{\mcitedefaultmidpunct}
{\mcitedefaultendpunct}{\mcitedefaultseppunct}\relax
\EndOfBibitem
\bibitem[Nguyen \latin{et~al.}(2018)Nguyen, Oh, Duong, Bang, Yoon, and
  Jeong]{nguyen_highly_2018}
Nguyen,~D.~A.; Oh,~H.~M.; Duong,~N.~T.; Bang,~S.; Yoon,~S.~J.; Jeong,~M.~S.
  Highly {Enhanced} {Photoresponsivity} of a {Monolayer} {WSe}$_2$ {Photodetector}
  with {Nitrogen}-{Doped} {Graphene} {Quantum} {Dots}. \emph{ACS Applied
  Materials \& Interfaces} \textbf{2018}, \emph{10}, 10322--10329\relax
\mciteBstWouldAddEndPuncttrue
\mciteSetBstMidEndSepPunct{\mcitedefaultmidpunct}
{\mcitedefaultendpunct}{\mcitedefaultseppunct}\relax
\EndOfBibitem
\bibitem[Radisavljevic and Kis(2013)Radisavljevic, and
  Kis]{radisavljevic_mobility_2013}
Radisavljevic,~B.; Kis,~A. Mobility Engineering and a Metal-Insulator
  Transition in Monolayer {MoS}$_{\textrm{2}}$. \emph{Nature Materials}
  \textbf{2013}, \emph{12}, 815--820\relax
\mciteBstWouldAddEndPuncttrue
\mciteSetBstMidEndSepPunct{\mcitedefaultmidpunct}
{\mcitedefaultendpunct}{\mcitedefaultseppunct}\relax
\EndOfBibitem
\bibitem[Radisavljevic \latin{et~al.}(2011)Radisavljevic, Radenovic, Brivio,
  Giacometti, and Kis]{radisavljevic_single-layer_2011}
Radisavljevic,~B.; Radenovic,~A.; Brivio,~J.; Giacometti,~V.; Kis,~A.
  Single-layer {MoS}$_{\textrm{2}}$ Transistors. \emph{Nature Nanotechnology}
  \textbf{2011}, \emph{6}, 147--150\relax
\mciteBstWouldAddEndPuncttrue
\mciteSetBstMidEndSepPunct{\mcitedefaultmidpunct}
{\mcitedefaultendpunct}{\mcitedefaultseppunct}\relax
\EndOfBibitem
\bibitem[Schwierz(2010)]{schwierz_graphene_2010}
Schwierz,~F. Graphene Transistors. \emph{Nature Nanotechnology} \textbf{2010},
  \emph{5}, 487--496\relax
\mciteBstWouldAddEndPuncttrue
\mciteSetBstMidEndSepPunct{\mcitedefaultmidpunct}
{\mcitedefaultendpunct}{\mcitedefaultseppunct}\relax
\EndOfBibitem
\bibitem[Xiao \latin{et~al.}(2012)Xiao, Liu, Feng, Xu, and
  Yao]{xiao_coupled_2012}
Xiao,~D.; Liu,~G.-B.; Feng,~W.; Xu,~X.; Yao,~W. Coupled {Spin} and {Valley}
  {Physics} in {Monolayers} of {MoS}$_{\textrm{2}}$
  and {Other} {Group}-{VI} {Dichalcogenides}. \emph{Physical Review Letters}
  \textbf{2012}, \emph{108}, 196802\relax
\mciteBstWouldAddEndPuncttrue
\mciteSetBstMidEndSepPunct{\mcitedefaultmidpunct}
{\mcitedefaultendpunct}{\mcitedefaultseppunct}\relax
\EndOfBibitem
\bibitem[Chen \latin{et~al.}(2017)Chen, Yan, Zhu, Yang, and
  Cui]{chen_optical_2017}
Chen,~X.; Yan,~T.; Zhu,~B.; Yang,~S.; Cui,~X. Optical {Control} of {Spin}
  {Polarization} in {Monolayer} {Transition} {Metal} {Dichalcogenides}.
  \emph{ACS Nano} \textbf{2017}, \emph{11}, 1581--1587\relax
\mciteBstWouldAddEndPuncttrue
\mciteSetBstMidEndSepPunct{\mcitedefaultmidpunct}
{\mcitedefaultendpunct}{\mcitedefaultseppunct}\relax
\EndOfBibitem
\bibitem[Mott(1938)]{mott_note_1938}
Mott,~N.~F. Note on the Contact Between a Metal and an insulator or
  Semiconductor. \emph{Mathematical Proceedings of the Cambridge Philosophical
  Society} \textbf{1938}, \emph{34}, 568--572\relax
\mciteBstWouldAddEndPuncttrue
\mciteSetBstMidEndSepPunct{\mcitedefaultmidpunct}
{\mcitedefaultendpunct}{\mcitedefaultseppunct}\relax
\EndOfBibitem
\bibitem[Ghiasi \latin{et~al.}(2018)Ghiasi, Quereda, and
  Wees]{ghiasi_bilayer_2018}
Ghiasi,~T.~S.; Quereda,~J.; Wees,~B. J.~v. Bilayer h-{BN} Barriers for
  Tunneling Contacts in Fully-Encapsulated Monolayer {MoSe$_2$} Field-Effect
  Transistors. \emph{2D Mater.} \textbf{2018}, \emph{6}, 015002\relax
\mciteBstWouldAddEndPuncttrue
\mciteSetBstMidEndSepPunct{\mcitedefaultmidpunct}
{\mcitedefaultendpunct}{\mcitedefaultseppunct}\relax
\EndOfBibitem
\bibitem[Cui \latin{et~al.}(2017)Cui, Shih, Jauregui, Chae, Kim, Li, Seo,
  Pistunova, Yin, Park, Choi, Lee, Watanabe, Taniguchi, Kim, Dean, and
  Hone]{cui_low-temperature_2017}
Cui,~X. \latin{et~al.}  Low-{Temperature} {Ohmic} {Contact} to {Monolayer}
  {MoS}$_2$ by Van der {Waals} {Bonded} {Co}/h-{BN} {Electrodes}. \emph{Nano
  Letters} \textbf{2017}, \emph{17}, 4781--4786\relax
\mciteBstWouldAddEndPuncttrue
\mciteSetBstMidEndSepPunct{\mcitedefaultmidpunct}
{\mcitedefaultendpunct}{\mcitedefaultseppunct}\relax
\EndOfBibitem
\bibitem[Chen \latin{et~al.}(2013)Chen, Odenthal, Swartz, Floyd, Wen, Luo, and
  Kawakami]{chen_control_2013}
Chen,~J.-R.; Odenthal,~P.~M.; Swartz,~A.~G.; Floyd,~G.~C.; Wen,~H.; Luo,~K.~Y.;
  Kawakami,~R.~K. Control of {Schottky} {Barriers} in {Single} {Layer} {MoS}$_2$
  {Transistors} with {Ferromagnetic} {Contacts}. \emph{Nano Letters}
  \textbf{2013}, \emph{13}, 3106--3110\relax
\mciteBstWouldAddEndPuncttrue
\mciteSetBstMidEndSepPunct{\mcitedefaultmidpunct}
{\mcitedefaultendpunct}{\mcitedefaultseppunct}\relax
\EndOfBibitem
\bibitem[Allain \latin{et~al.}(2015)Allain, Kang, Banerjee, and
  Kis]{allain_electrical_2015}
Allain,~A.; Kang,~J.; Banerjee,~K.; Kis,~A. Electrical Contacts to
  Two-Dimensional Semiconductors. \emph{Nature Materials} \textbf{2015},
  \emph{14}, 1195--1205\relax
\mciteBstWouldAddEndPuncttrue
\mciteSetBstMidEndSepPunct{\mcitedefaultmidpunct}
{\mcitedefaultendpunct}{\mcitedefaultseppunct}\relax
\EndOfBibitem
\bibitem[Zhou \latin{et~al.}(2019)Zhou, Taguchi, Kawaguchi, Tanaka, and
  Law]{zhou2019spin}
Zhou,~B.~T.; Taguchi,~K.; Kawaguchi,~Y.; Tanaka,~Y.; Law,~K. Spin-Orbit
  Coupling Induced Valley Hall Effects in Transition-Metal Dichalcogenides.
  \emph{Communications Physics} \textbf{2019}, \emph{2}, 26\relax
\mciteBstWouldAddEndPuncttrue
\mciteSetBstMidEndSepPunct{\mcitedefaultmidpunct}
{\mcitedefaultendpunct}{\mcitedefaultseppunct}\relax
\EndOfBibitem
\bibitem[Anwar \latin{et~al.}(1999)Anwar, Nabet, Culp, and
  Castro]{anwar_effects_1999}
Anwar,~A.; Nabet,~B.; Culp,~J.; Castro,~F. Effects of Electron Confinement on
  Thermionic Emission Current in a Modulation Doped Heterostructure.
  \emph{Journal of Applied Physics} \textbf{1999}, \emph{85}, 2663--2666\relax
\mciteBstWouldAddEndPuncttrue
\mciteSetBstMidEndSepPunct{\mcitedefaultmidpunct}
{\mcitedefaultendpunct}{\mcitedefaultseppunct}\relax
\EndOfBibitem
\bibitem[Bhuiyan \latin{et~al.}(1988)Bhuiyan, Martinez, and
  Esteve]{bhuiyan_new_1988}
Bhuiyan,~A.~S.; Martinez,~A.; Esteve,~D. A New {Richardson} Plot for Non-Ideal
  Schottky Diodes. \emph{Thin Solid Films} \textbf{1988}, \emph{161},
  93--100\relax
\mciteBstWouldAddEndPuncttrue
\mciteSetBstMidEndSepPunct{\mcitedefaultmidpunct}
{\mcitedefaultendpunct}{\mcitedefaultseppunct}\relax
\EndOfBibitem
\bibitem[Jariwala \latin{et~al.}(2013)Jariwala, Sangwan, Late, Johns, Dravid,
  Marks, Lauhon, and Hersam]{jariwala_band-like_2013}
Jariwala,~D.; Sangwan,~V.~K.; Late,~D.~J.; Johns,~J.~E.; Dravid,~V.~P.;
  Marks,~T.~J.; Lauhon,~L.~J.; Hersam,~M.~C. Band-like Transport in High
  Mobility Unencapsulated Single-Layer {MoS}$_2$ Transistors. \emph{Applied
  Physics Letters} \textbf{2013}, \emph{102}, 173107\relax
\mciteBstWouldAddEndPuncttrue
\mciteSetBstMidEndSepPunct{\mcitedefaultmidpunct}
{\mcitedefaultendpunct}{\mcitedefaultseppunct}\relax
\EndOfBibitem
\bibitem[Yamaguchi \latin{et~al.}(2014)Yamaguchi, Moriya, Inoue, Morikawa,
  Masubuchi, Watanabe, Taniguchi, and Machida]{yamaguchi2014tunneling}
Yamaguchi,~T.; Moriya,~R.; Inoue,~Y.; Morikawa,~S.; Masubuchi,~S.;
  Watanabe,~K.; Taniguchi,~T.; Machida,~T. Tunneling Transport in a Few
  Monolayer-Thick WS$_2$/Graphene Heterojunction. \emph{Applied Physics Letters}
  \textbf{2014}, \emph{105}, 223109\relax
\mciteBstWouldAddEndPuncttrue
\mciteSetBstMidEndSepPunct{\mcitedefaultmidpunct}
{\mcitedefaultendpunct}{\mcitedefaultseppunct}\relax
\EndOfBibitem
\bibitem[Moriya \latin{et~al.}(2014)Moriya, Yamaguchi, Inoue, Morikawa, Sata,
  Masubuchi, and Machida]{moriya2014large}
Moriya,~R.; Yamaguchi,~T.; Inoue,~Y.; Morikawa,~S.; Sata,~Y.; Masubuchi,~S.;
  Machida,~T. Large Current Modulation in Exfoliated-Graphene/MoS$_2$/Metal
  Vertical Heterostructures. \emph{Applied Physics Letters} \textbf{2014},
  \emph{105}, 083119\relax
\mciteBstWouldAddEndPuncttrue
\mciteSetBstMidEndSepPunct{\mcitedefaultmidpunct}
{\mcitedefaultendpunct}{\mcitedefaultseppunct}\relax
\EndOfBibitem
\bibitem[Omar and van Wees(2017)Omar, and van Wees]{omar2017graphene}
Omar,~S.; van Wees,~B.~J. Graphene-WS$_2$ Heterostructures for Tunable Spin
  Injection and Spin transport. \emph{Physical Review B} \textbf{2017},
  \emph{95}, 081404\relax
\mciteBstWouldAddEndPuncttrue
\mciteSetBstMidEndSepPunct{\mcitedefaultmidpunct}
{\mcitedefaultendpunct}{\mcitedefaultseppunct}\relax
\EndOfBibitem
\bibitem[Liang \latin{et~al.}(2017)Liang, Yang, Renucci, Tao, Laczkowski,
  Mc-Murtry, Wang, Marie, George, Petit-Watelot, \latin{et~al.}
  others]{liang2017electrical}
Liang,~S.; Yang,~H.; Renucci,~P.; Tao,~B.; Laczkowski,~P.; Mc-Murtry,~S.;
  Wang,~G.; Marie,~X.; George,~J.-M.; Petit-Watelot,~S., \latin{et~al.}
  Electrical Spin Injection and Detection in Molybdenum Disulfide Multilayer
  Channel. \emph{Nature communications} \textbf{2017}, \emph{8}, 1--9\relax
\mciteBstWouldAddEndPuncttrue
\mciteSetBstMidEndSepPunct{\mcitedefaultmidpunct}
{\mcitedefaultendpunct}{\mcitedefaultseppunct}\relax
\EndOfBibitem
\bibitem[Liu \latin{et~al.}(2013)Liu, Cao, Kang, and Banerjee]{liu2013high}
Liu,~W.; Cao,~W.; Kang,~J.; Banerjee,~K. High-Performance
  Field-Effect-Transistors on Monolayer-WSe$_2$. \emph{ECS Transactions}
  \textbf{2013}, \emph{58}, 281--285\relax
\mciteBstWouldAddEndPuncttrue
\mciteSetBstMidEndSepPunct{\mcitedefaultmidpunct}
{\mcitedefaultendpunct}{\mcitedefaultseppunct}\relax
\EndOfBibitem
\bibitem[Neamen(2012)]{donaldjneamen2012}
Neamen,~D.~J. Semiconductor Physics And Devices. \emph{McGraw-Hill Education -
  Europe} \textbf{2012}, 336--337\relax
\mciteBstWouldAddEndPuncttrue
\mciteSetBstMidEndSepPunct{\mcitedefaultmidpunct}
{\mcitedefaultendpunct}{\mcitedefaultseppunct}\relax
\EndOfBibitem
\bibitem[Das \latin{et~al.}(2012)Das, Chen, Penumatcha, and
  Appenzeller]{das2012high}
Das,~S.; Chen,~H.-Y.; Penumatcha,~A.~V.; Appenzeller,~J. High Performance
  Multilayer MoS$_2$ Transistors with Scandium Contacts. \emph{Nano letters}
  \textbf{2012}, \emph{13}, 100--105\relax
\mciteBstWouldAddEndPuncttrue
\mciteSetBstMidEndSepPunct{\mcitedefaultmidpunct}
{\mcitedefaultendpunct}{\mcitedefaultseppunct}\relax
\EndOfBibitem
\bibitem[Hughes \latin{et~al.}(1982)Hughes, McKinley, Williams, and
  McGovern]{hughes_metal-gallium_1982}
Hughes,~G.~J.; McKinley,~A.; Williams,~R.~H.; McGovern,~I.~T. Metal-Gallium
  Selenide Interfaces-Observation of the True {Schottky} Limit. \emph{Journal
  of Physics C: Solid State Physics} \textbf{1982}, \emph{15}, L159--L164\relax
\mciteBstWouldAddEndPuncttrue
\mciteSetBstMidEndSepPunct{\mcitedefaultmidpunct}
{\mcitedefaultendpunct}{\mcitedefaultseppunct}\relax
\EndOfBibitem
\bibitem[LaGasse \latin{et~al.}()LaGasse, Dhakras, Watanabe, Taniguchi, and
  Lee]{lagasse_gate-tunable_nodate}
LaGasse,~S.~W.; Dhakras,~P.; Watanabe,~K.; Taniguchi,~T.; Lee,~J.~U.
  Gate-{Tunable} {Graphene}–{WSe}$_2$ {Heterojunctions} at the
  {Schottky}–{Mott} {Limit}. \emph{Advanced Materials} 
  \textbf{2019}, \emph{31},
  1901392\relax
\mciteBstWouldAddEndPuncttrue
\mciteSetBstMidEndSepPunct{\mcitedefaultmidpunct}
{\mcitedefaultendpunct}{\mcitedefaultseppunct}\relax
\EndOfBibitem
\bibitem[Cowley and Sze(1965)Cowley, and Sze]{cowley1965surface}
Cowley,~A.; Sze,~S. Surface States and Barrier Height of Metal-Semiconductor
  Systems. \emph{Journal of Applied Physics} \textbf{1965}, \emph{36},
  3212--3220\relax
\mciteBstWouldAddEndPuncttrue
\mciteSetBstMidEndSepPunct{\mcitedefaultmidpunct}
{\mcitedefaultendpunct}{\mcitedefaultseppunct}\relax
\EndOfBibitem
\bibitem[Stanford \latin{et~al.}(2017)Stanford, Pudasaini, Gallmeier, Cross,
  Liang, Oyedele, Duscher, Mahjouri-Samani, Wang, and Xiao]{stanford2017high}
Stanford,~M.~G.; Pudasaini,~P.~R.; Gallmeier,~E.~T.; Cross,~N.; Liang,~L.;
  Oyedele,~A.; Duscher,~G.; Mahjouri-Samani,~M.; Wang,~K.; Xiao,~K. High
  Conduction Hopping Behavior Induced in Transition Metal Dichalcogenides by
  Percolating Defect Networks: Toward Atomically Thin Circuits. \emph{Advanced
  Functional Materials} \textbf{2017}, \emph{27}, 1702829\relax
\mciteBstWouldAddEndPuncttrue
\mciteSetBstMidEndSepPunct{\mcitedefaultmidpunct}
{\mcitedefaultendpunct}{\mcitedefaultseppunct}\relax
\EndOfBibitem
\bibitem[Hamilton(1972)]{doi:10.1080/14786437208226975}
Hamilton,~E.~M. Variable Range Hopping in a Non-Uniform Density of States.
  \emph{The Philosophical Magazine: A Journal of Theoretical Experimental and
  Applied Physics} \textbf{1972}, \emph{26}, 1043--1045\relax
\mciteBstWouldAddEndPuncttrue
\mciteSetBstMidEndSepPunct{\mcitedefaultmidpunct}
{\mcitedefaultendpunct}{\mcitedefaultseppunct}\relax
\EndOfBibitem
\bibitem[Qiu \latin{et~al.}(2013)Qiu, Xu, Wang, Ren, Nan, Ni, Chen, Yuan, Miao,
  Song, Long, Shi, Sun, Wang, and Wang]{qiu_hopping_2013}
Qiu,~H.; Xu,~T.; Wang,~Z.; Ren,~W.; Nan,~H.; Ni,~Z.; Chen,~Q.; Yuan,~S.;
  Miao,~F.; Song,~F.; Long,~G.; Shi,~Y.; Sun,~L.; Wang,~J.; Wang,~X. Hopping
  Transport Through Defect-Induced Localized States in Molybdenum Disulphide.
  \emph{Nature Communications} \textbf{2013}, \emph{4}, 2642\relax
\mciteBstWouldAddEndPuncttrue
\mciteSetBstMidEndSepPunct{\mcitedefaultmidpunct}
{\mcitedefaultendpunct}{\mcitedefaultseppunct}\relax
\EndOfBibitem
\bibitem[Han \latin{et~al.}(2010)Han, Brant, and Kim]{han_electron_2010}
Han,~M.~Y.; Brant,~J.~C.; Kim,~P. Electron {Transport} in {Disordered}
  {Graphene} {Nanoribbons}. \emph{Physical Review Letters} \textbf{2010},
  \emph{104}, 056801\relax
\mciteBstWouldAddEndPuncttrue
\mciteSetBstMidEndSepPunct{\mcitedefaultmidpunct}
{\mcitedefaultendpunct}{\mcitedefaultseppunct}\relax
\EndOfBibitem
\bibitem[Yu and Song(2001)Yu, and Song]{yu_variable_2001}
Yu,~Z.~G.; Song,~X. Variable {Range} {Hopping} and {Electrical} {Conductivity}
  along the {DNA} {Double} {Helix}. \emph{Physical Review Letters}
  \textbf{2001}, \emph{86}, 6018--6021\relax
\mciteBstWouldAddEndPuncttrue
\mciteSetBstMidEndSepPunct{\mcitedefaultmidpunct}
{\mcitedefaultendpunct}{\mcitedefaultseppunct}\relax
\EndOfBibitem
\bibitem[Lu \latin{et~al.}(2014)Lu, Lacour, Lengaigne, Le~Gall, Suire,
  Montaigne, Hehn, and Wu]{lu_electrical_2014}
Lu,~Y.; Lacour,~D.; Lengaigne,~G.; Le~Gall,~S.; Suire,~S.; Montaigne,~F.;
  Hehn,~M.; Wu,~M.~W. Electrical Control of Interfacial Trapping for Magnetic
  Tunnel Transistor on Silicon. \emph{Applied Physics Letters} \textbf{2014},
  \emph{104}, 042408\relax
\mciteBstWouldAddEndPuncttrue
\mciteSetBstMidEndSepPunct{\mcitedefaultmidpunct}
{\mcitedefaultendpunct}{\mcitedefaultseppunct}\relax
\EndOfBibitem
\bibitem[Lu \latin{et~al.}(2009)Lu, Tran, Jaffrès, Seneor, Deranlot, Petroff,
  George, Lépine, Ababou, and Jézéquel]{lu_spin-polarized_2009}
Lu,~Y.; Tran,~M.; Jaffrès,~H.; Seneor,~P.; Deranlot,~C.; Petroff,~F.;
  George,~J.-M.; Lépine,~B.; Ababou,~S.; Jézéquel,~G. Spin-{Polarized}
  {Inelastic} {Tunneling} through {Insulating} {Barriers}. \emph{Physical
  Review Letters} \textbf{2009}, \emph{102}, 176801\relax
\mciteBstWouldAddEndPuncttrue
\mciteSetBstMidEndSepPunct{\mcitedefaultmidpunct}
{\mcitedefaultendpunct}{\mcitedefaultseppunct}\relax
\EndOfBibitem

\end{mcitethebibliography}
\providecommand{\latin}[1]{#1}
\makeatletter
\providecommand{\doi}
  {\begingroup\let\do\@makeother\dospecials
  \catcode`\{=1 \catcode`\}=2 \doi@aux}
\providecommand{\doi@aux}[1]{\endgroup\texttt{#1}}
\makeatother
\providecommand*\mcitethebibliography{\thebibliography}
\csname @ifundefined\endcsname{endmcitethebibliography}
  {\let\endmcitethebibliography\endthebibliography}{}

\end{document}